\documentclass[preprint,12pt]{elsarticle}

\usepackage{subfigure}
\usepackage{caption}
\usepackage{amssymb}
\journal{Elsevier}
\begin{document}

\begin{frontmatter}



\title{Fractional derivative defined by non-singular kernels to capture anomalous relaxation and diffusion}


\author{HongGuang Sun$^{a,b}$, Xiaoxiao Hao$^{a}$, Yong Zhang$^{b}$, Dumitru Baleanu$^{c,d}$}

\address{a. Institute of Soft Matter Mechanics, Department of Engineering
Mechanics, Hohai University, 1 XiKang Road, Nanjing, Jiangsu
210098, China\\
b. Department of Geological Sciences, University of Alabama, Tuscaloosa, AL 35487, USA\\
c. Department of Mathematics, Cankaya University, Ankara, Turkey\\
d. Institute of Space Sciences, Magurele-Bucharest, Romania}

\begin{abstract}
Anomalous relaxation and diffusion processes have been widely characterized by fractional derivative models, where the definition of the fractional-order derivative remains a historical debate due to the singular memory kernel that challenges numerical calculations. This study first explores physical properties of relaxation and diffusion models where the fractional derivative was defined recently using an exponential kernel. Analytical analysis shows that the fractional derivative model with an exponential kernel cannot characterize non-exponential dynamics well-documented in anomalous relaxation and diffusion. A legitimate extension of the previous fractional derivative is then proposed by replacing the exponential kernel with a stretched exponential kernel. Numerical tests show that the fractional derivative model with the stretched exponential kernel can describe a much wider range of anomalous diffusion than the exponential kernel, implying the potential applicability of the new fractional derivative in quantifying real-world, anomalous relaxation and diffusion processes.

\end{abstract}

\begin{keyword}
Anomalous relaxation and diffusion, Non-singular kernel, Stretched exponential function kernel, Memory characterization, Mean squared displacement%% keywords here, in the form: keyword \sep keyword

%\PACS 66.10.C-, 05.10.Gg, 02.60.Cb % PACS, the Physics and Astronomy
                             % Classification Scheme. \sep code


\end{keyword}

\end{frontmatter}


\section{Introduction}
In physics, processes involving the basic phenomena of relaxation and diffusion are of great relevance \cite{Mainardi1996,Kakalios1987}.  Many non-equilibrium physical systems and processes can be characterized by relaxation and diffusion equations \cite{Borman2016,Jeon2013}. However, anomalous relaxation and diffusion processes in various complex systems do not usually follow classical exponential decay or Gaussian statistics, but rather exhibit different relaxation or diffusion patterns. The traditional relaxation equation or Fick's second law therefore fails to describe the related behaviors \cite{Meerschaert2015,Bouchaud1990,Metzler2014}, motivating the development of new mathematical or physical equation models \cite{Li2013,Ngai2011,Metzler2000}. Fractional derivatives, which can contain the historical memory and global correlation information of complex physical systems, have been considered a successful tool to characterize anomalous dynamics in various physics and engineering processes \cite{Magin2013,Yu2015,Zeng2013,Tomovski2012}. Hereby, the fractional-derivative based relaxation equation and the corresponding diffusion equation models (which govern many physical phenomena such as heat, mass, or electron transfer; pollutants or liquid transport through porous media; amorphous semiconductors; colloid or proteins moving in biosystems or even in ecosystems) have been widely investigated in multiple disciplines including physics, mathematics, mechanics, and hydrology and control theory, among many others \cite{Metzler2014,Yu2015,Wu2015,Uchaikin2012}.

From the viewpoint of mathematics, the reason that the fractional-derivative model can characterize the system memory and global correlation well is due to its standard power-law memory kernel used to define the fractional derivative. However, the singularity of the power-law kernel is the main problem in numerical computation and applications of fractional-order partial differential equations (PDEs). To solve this problem, a new definition of the fractional derivative with an exponential kernel was recently proposed and further investigated from mathematical and physical aspects by Caputo and Fabrizi \cite{Caputo2015}. The main advantage of the new definition is that the singular power-law kernel is now replaced by a non-singular exponential kernel, which is easier to use in theoretical analysis, numerical calculations and real-world applications. The mathematical properties of the new definition such as the existence of solution \cite{Atangana2015} and Laplace transform \cite{Losada2015}  have been investigated. Potential application of fractional derivative models with an exponential kernel was recently investigated for electromagnetism \cite{Caputo2016}, diffusion \cite{Aguilar2016}, and heat transfer \cite{Atangana2016a}.

Physical aspects of fractional derivative models where the memory kernel is an exponential function, however, have not been well documented, motivating this study. In the following we first analyze the analytical solution and the mean squared displacement for relaxation and diffusion governed by fractional derivative models with an exponential kernel. Results show that fractional derivative models with an exponential kernel cannot efficiently characterize the non-exponential nature of anomalous relaxation and diffusion. Hereby, a new notion of a fractional derivative where the memory kernel takes the form of a stretched exponential function is proposed to overcome the limitations of the existing exponential kernel. For the sake of simplicity, but without the loss of generality, we consider one dimensional relaxation and diffusion equations with a temporal fractional derivative.

\section{Fractional derivative with non-singular kernels}

The standard Caputo type fractional derivative contains a singular power-law kernel \cite{Caputo1965,Baleanu2006}:
\begin{eqnarray}
^{C}D_t^{\alpha}f(t)=\frac{1}{\Gamma(1-\alpha)}\int_{a}^t\frac{f^{'}(\tau)d\tau}{(t-\tau)^{\alpha}},\quad
0<\alpha< 1,\label{eq2-1}
\end{eqnarray}
in which $\alpha$ is the order of the fractional derivative, $a\in (-\infty, t)$ is the initial time in the history, and $\Gamma(\cdot)$ denotes the Gamma function.

To overcome the complexity caused by using a singular memory kernel $(t-\tau)^{-\alpha}$, Caputo and Fabrizio \cite{Caputo2015} recently proposed a new definition of a factional derivative using the non-singular exponential function:
\begin{eqnarray}
\begin{array}{c}
\displaystyle{^{C1}D^\alpha f(t)=\frac{M(\alpha)}{(1-\alpha)}\int_a^t f'(\tau) exp[-\frac{\alpha (t-\tau)}{1-\alpha}] d \tau,}
\end{array}
\label{eq2-2}
\end{eqnarray}
where the super-script ``$C1$" denotes the Caputo and Fabrizio type fractional derivative with an exponential memory, and $M(\alpha)$ is a normalization function such that $M(0)=M(1)=1$. For simplicity, we choose $M(\alpha)=1$ in the following sections. For $\beta>1$, we select $\beta =n+\alpha$, where $\alpha\in [0, 1]$ and $n$ is an integer. The $\beta$-order Caputo and Fabrizio type fractional derivative can be expressed as:
 \begin{eqnarray}
\begin{array}{c}
\displaystyle{^{C1}D^\beta f(t)=^{C1}D^\alpha f^{(n)}(t)=\frac{M(\alpha)}{(1-\alpha)}\int_a^t f^{n+1}(\tau) exp[-\frac{\alpha (t-\tau)}{1-\alpha}] d \tau.}
\end{array}
\label{eq7}
\end{eqnarray}

Caputo and Fabrizio \cite{Caputo2016} also investigated the potential expression of fractional derivative with a Gaussian-function kernel for gradient and Laplacian operators:
\begin{eqnarray}
\begin{array}{c}
\displaystyle{^{C2}D^\alpha f(t)=\frac{1+\alpha^2}{\sqrt{\pi^\alpha(1-\alpha)}}\int_a^t f'(\tau) exp[-\frac{\alpha (t-\tau)^2}{1-\alpha}] d \tau,}
\end{array}
\label{eq2-3}
\end{eqnarray}
which will be briefly analyzed in the following.

%If we consider the influence of initial condition, the corresponding definition with exponential kernel should be expressed as
%\begin{eqnarray}
%\begin{array}{c}
%\displaystyle{^{C1}D^\alpha f(t)=\frac{1}{(1-\alpha)}\int_a^t f'(\tau) exp[-\frac{\alpha (t-\tau)}{1-\alpha}] d \tau-f(0)exp[-\frac{\alpha t}{1-\alpha}],}
%\end{array}
%\label{eq2-4}
%\end{eqnarray}
In addition, another interesting definition of a fractional derivative with a Mittag-Leffler function kernel has been proposed and further discussed by Atangana and Baleanu \cite{Atangana2016a,Atangana2016b}. That definition was not considered in this study due to the computational burden of the Mittag-Leffler function and the corresponding fractional derivative.

\section{Power-law and exponential function kernels in characterizing relaxation and diffusion processes}
Here we explore the dynamics of relaxation and diffusion processes characterized by fractional derivative models with known memory kernels.

\subsection{Fractional relaxation equation model}
The following fractional-order relaxation equation is a simple and typical model used to describe various physical processes:
\begin{eqnarray}
\left\{
\begin{array}{l}
\displaystyle{\frac{d^\alpha u(t)}{d t^\alpha}=- \lambda u(t), 0<\alpha \leq 1,} \\
\displaystyle{u(t=0)=U_0.}
\end{array}\right.\;
\label{eq3-1}
\end{eqnarray}
%\begin{eqnarray}
%\begin{array}{c}
%\displaystyle{\frac{d^\alpha u(t)}{d t^\alpha}=- \lambda u(t), 0<\alpha \leq 1,}
%\displaystyle{u(t=0)=U_0.}
%\end{array}
%\label{eq3-1}
%\end{eqnarray}
The variable $u(t)$ can represent concentration in solute transport \cite{Metzler2000,Chechkin2008,Zhang2009}, the relaxation modulus in viscoelasticity \cite{Mainardi2010}, or input in control systems \cite{Sheng2011}. It is well known that this equation with an integer order ($\alpha=1$) can characterize the phenomena of exponential relaxation in physics.

\textbf{(a) Fractional derivative relaxation equation with a power-law kernel}

First we consider a fractional derivative with the traditional power-law function kernel. The Laplace transform of Eq. (\ref{eq3-1}) can be stated as:
\begin{eqnarray}
\begin{array}{c}
\displaystyle{s \overline{u}(s)-U_0 s^{\alpha-1}=- \lambda \overline{u}(s), 0<\alpha \leq 1.}
\end{array}
\label{eq3-2}
\end{eqnarray}
Inverse Laplace transform of Eq. (\ref{eq3-2}) provides the analytical solution of Eq. (\ref{eq3-1}) \cite{Mainardi2010}:
\begin{eqnarray}
\begin{array}{c}
\displaystyle{u(t)=U_0 E_\alpha (-\lambda t^\alpha),}
\end{array}
\label{eq3-3}
\end{eqnarray}
where $E_\alpha(\cdot)$ denotes the Mittag-Leffler function\cite{Metzler2000,Sun2010}.

\textbf{(b) Fractional derivative relaxation equation with an exponential kernel}

To obtain the analytical solution of Eq.(\ref{eq3-1}) with an exponential kernel, we first use the Laplace transform of the fractional derivative with an exponential kernel Eq.(\ref{eq2-2}) \cite{Caputo2015,Losada2015}
\begin{eqnarray}
\begin{array}{c}
\displaystyle{L[^{C1}D^\alpha u(t)] (s)=\frac{1}{1-\alpha} L(u'(t)) L[exp(\frac{-\alpha t}{1-\alpha})]}\\
\displaystyle{=\frac{s u(s)-U_0}{s(1-\alpha)+\alpha}.}
\end{array}
\label{eq3-3}
\end{eqnarray}
Then one can get the expression of Eq. (\ref{eq3-1}) after Laplace transform
\begin{eqnarray}
\begin{array}{c}
\displaystyle{\frac{s u(s)-U_0}{s(1-\alpha)+\alpha}=-\lambda u(s).}\\
\end{array}
\label{eq3-4}
\end{eqnarray}
Finally, the analytical solution of Eq. (\ref{eq3-1}) can be obtained via an inverse Laplace transform with the initial condition defined in (\ref{eq3-1}):
\begin{eqnarray}
\begin{array}{c}
\displaystyle{u(t)=U_0 e^{-\frac{\lambda \alpha t}{1+\lambda-\alpha \lambda}}.}\\
%\displaystyle{u(t)=\frac{U_0}{1+\lambda-\alpha \lambda}e^{-\frac{\lambda \alpha t}{1+\lambda-\alpha \lambda}}.}\\
\end{array}
\label{eq3-5}
\end{eqnarray}

\textbf{Remarks.} A well-known and interesting feature of anomalous relaxation in complex systems is their non-exponential nature \cite{Stanislavsky2015}. The solution of fractional derivative models with a power-law function kernel exhibits power-law decay when $t\rightarrow \infty$, while the solution of that with an exponential function kernel shows an exponential decay similar to the classical integer-order model. In fact, the integer-order relaxation equation with a relaxation coefficient $\lambda \alpha/ (1+\lambda-\alpha \lambda)$ gives the same analytical solution (\ref{eq3-3}), from mathematical viewpoint.

\subsection{Mean squared displacement of the fractional derivative diffusion model}
Mean squared displacement (MSD) offers important information of diffusive motion in physical and biophysical systems, and serves an important criterion in anomalous diffusion \cite{Metzler2000,Kepten2015,Sun2010}. To investigate the physical behavior of fractional derivative models with an exponential function kernel, we consider the following time fractional diffusion equation:
\begin{eqnarray}
\left\{
\begin{array}{c}
\displaystyle{\frac{\partial^\alpha c(x,t)}{\partial t^\alpha}=K \frac{\partial^2
c(x,t)}{\partial x^2}, \,
 -\infty< x< +\infty,\, t>0,}\\
\displaystyle{c(x,0)=\delta(x),\,c(\pm\infty,t)=0,\,\frac{\partial
c(\pm\infty,t)}{\partial x}=0,}
\end{array}\right.\;
\label{eq4-1}
\end{eqnarray}
where $\delta(x)$ is the Dirac delta function, $\alpha \in (0, 1]$
denotes the order of the time fractional derivative, $K$ is a
generalized diffusion coefficient, $c(x,t)$ represents the target concentration.

Previous work found that the MSD of model (\ref{eq4-1}) with the traditional Caputo fractional derivative expressed by Eq. (\ref{eq2-1}) can be stated as \cite{Metzler2000}

\begin{equation}
\displaystyle{ \langle
x^2(t)\rangle=\frac{2K t^{\alpha}}{\Gamma(\alpha+1)}.}\label{eq4-2}
\end{equation}

We then derive the MSD for the fractional derivative model Eq.(\ref{eq4-1}) with an exponential kernel. At the first step, we
take integration of both sides of Eq. (\ref{eq4-1}) with $\int_{-\infty}^\infty
x^2 {\rm d}x$, which leads to
\begin{equation}
\displaystyle{\int_{-\infty}^\infty x^2\,{^{C1}D^{\alpha}}
c(x,t){\rm d}x =\int_{-\infty}^\infty x^2 K \frac{\partial^2
c(x,t)}{\partial x^2} {\rm d} x.}\label{eq4-3}
\end{equation}
It can be written as the following form by changing the order of integral and differentiation:
\begin{equation}
\displaystyle{^{C1}D_{0_+}^{\alpha} \langle x^2(t)\rangle
=\int_{-\infty}^\infty x^2 K \frac{\partial^2 c(x,t)}{\partial x^2}
{\rm d} x.}\label{eq4-4}
\end{equation}
Clearly, Eq. (\ref{eq4-4}) can be simplified into the following form based on the boundary conditions of model Eq. (\ref{eq4-1})
\begin{equation}
\displaystyle{^{C1} D_{0_+}^{\alpha} \langle x^2(t)\rangle =2
K.}\label{eq4-5}
\end{equation}
The final expression of the MSD can be obtained by applying the Laplace transform of Eq. (\ref{eq4-5}) and then taking inverse Laplace transform:
\begin{equation}
\displaystyle{ \langle
x^2(t)\rangle=2K \alpha t.}\label{eq4-6}
\end{equation}

\textbf{Remarks.} It is obvious that the fractional derivative diffusion equation model with an exponential kernel reduces to normal diffusion based on the criterion of MSD.

\section{Fractional derivative model with the memory kernel as a stretched exponential function}

Fractional derivative models with exponential kernels have obvious limitations when characterizing anomalous relaxation and diffusion of a non-exponential nature. Meanwhile, previous literature has also shown that the memory kernel appearing in the generalized diffusion equation has various potential forms which can describe more experimental phenomena \cite{Kakalios1987,Sandev2015,Sokolov2005}. Hereby, we consider a legitimate extension of the fractional derivative by replacing the exponential function kernel by a stretched exponential function kernel:
\begin{eqnarray}
\begin{array}{c}
\displaystyle{^{SC}D^\alpha f(t)=\frac{M(\alpha)}{(1-\alpha)^{1/\alpha}}\int_a^t f'(\tau) exp[-\frac{\alpha (t-\tau)^{\alpha} }{1-\alpha}] d \tau,\,\,0<\alpha<1,}
\end{array}
\label{eq6}
\end{eqnarray}
in which the super-script ``SC" represents the stretched exponential type Caputo definition, and $M(\alpha)$ is a normalization function. For simplicity, here we take $M(\alpha)=1/\Gamma(1+\alpha)$.

For $\beta>1$, we can let $\beta =n+\alpha$, where $\alpha\in (0, 1]$ and $n$ is an integer.  The $\beta$-order fractional derivative can be expressed as:
\begin{eqnarray}
\begin{array}{c}
\displaystyle{^{SC}D^\beta f(t)=^{SC}D^\alpha f^{(n)}(t)=\frac{M(\alpha)}{(1-\alpha)^{1/\alpha}}\int_a^t f^{n+1}(\tau) exp[-\frac{\alpha (t-\tau)^\alpha }{1-\alpha}] d \tau.}
\end{array}
\label{eq7}
\end{eqnarray}

\begin{figure}
\begin{center}
\setlength{\unitlength}{0.012500in}%
{\includegraphics[width=1.0\textwidth]{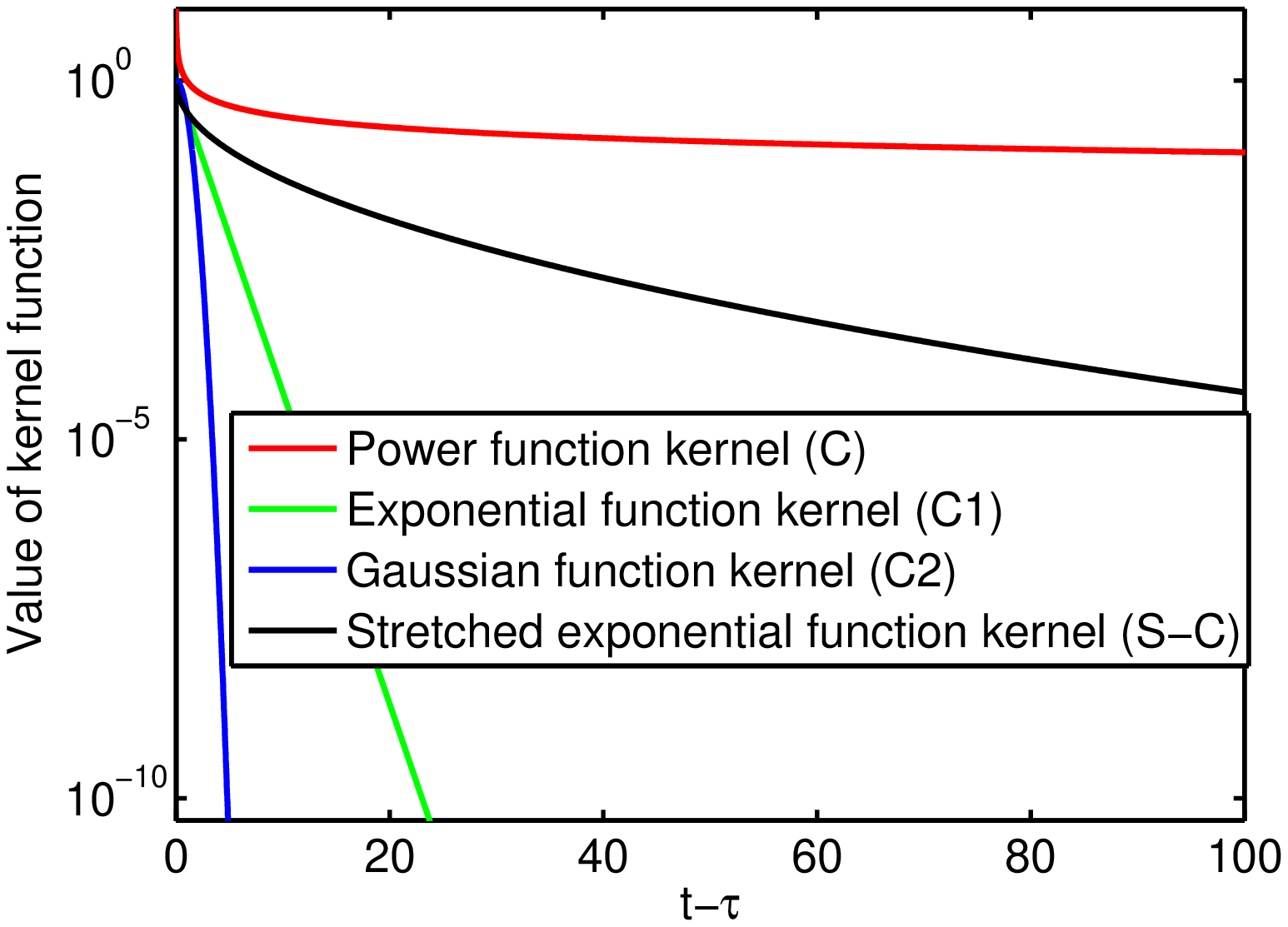}}\\
\end{center}
 \caption{Memory property represented by using different kernels with $\alpha=0.5$: the power-law function kernel is $(t-\tau)^{-\alpha}$, the exponential function kernel $exp(-\frac{\alpha*(t-\tau)}{1-\alpha})$, the Gaussian function kernel $exp(-\frac{\alpha*(t-\tau)^2}{1-\alpha})$, and the stretched exponential function kernel $exp(-\frac{\alpha*(t-\tau)^\alpha}{1-\alpha})$.}
 \label{figure1}
\end{figure}

Since the above fractional derivatives differ only in the functional form of the memory kernel, we first compare the memory behavior of different kernels. Figure \ref{figure1} shows that the power-law kernel represents strong memory in long time, compared to the exponential, stretched exponential, and Gaussian function kernels. Although the fractional derivative defined by the exponential kernel represents a process with memory, the long-time memory is very weak, which may explain why the solutions of the corresponding relaxation and diffusion equations are similar to classical integer-order derivative models. While the stretched exponential function kernel offers a relatively strong memory in long time, the memory property in small and long-time histories can be balanced in a wide range by changing the derivative order $\alpha$. Therefore, from this viewpoint, the stretched exponential function kernel is a better choice to characterize the memory (time) or nonlocal (space) process when using the fractional derivative with non-singular kernels.
% Figure 1

Next, we investigate the physical behavior of fractional derivative relaxation and diffusion equations with a stretched exponential kernel, where the analytical solutions are not available.  We use a finite difference numerical scheme to obtain accurate solutions of considered physical quantities \cite{Diethelm2002,Zhao2015,Wang2010}.

\begin{figure}[t]
%\begin{figure}
\begin{center}
\setlength{\unitlength}{0.012500in}%
{\includegraphics[width=1.0\textwidth]{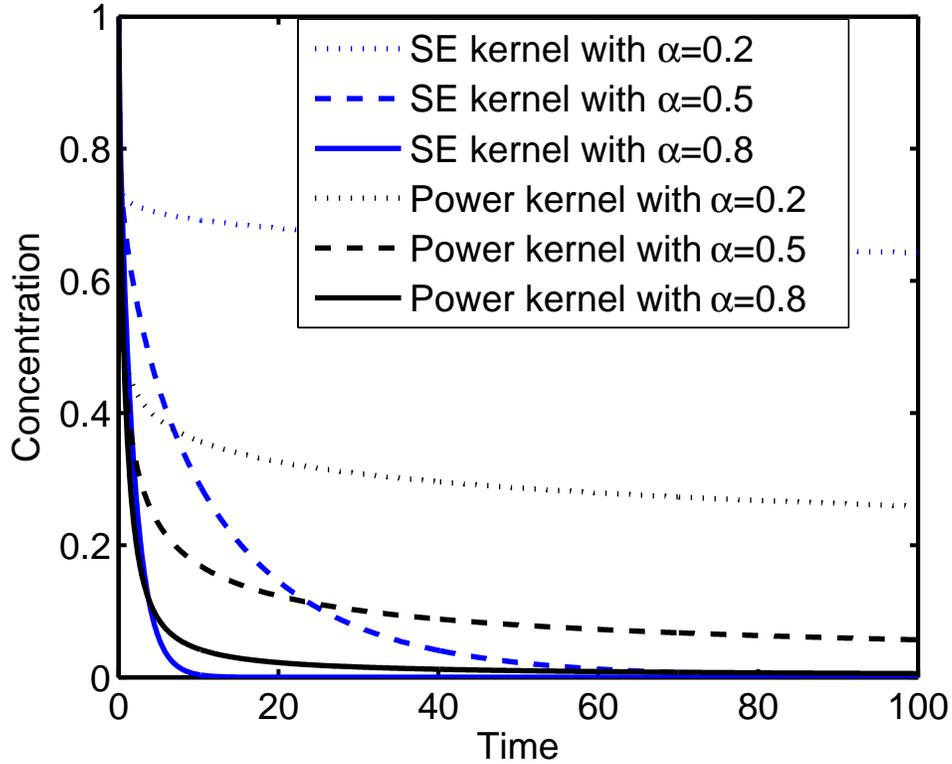}}\\
\end{center}
 \caption{Comparison of fractional derivative relaxation equations with a power-law kernel and a stretched exponential kernel. Here the legend ``SE" represents stretched exponential. The blue lines represent concentration relaxation curves drawn by using the stretched exponential function kernel with $\alpha=0.2, 0.5$ and $0.8$ (from top to bottom). The black lines denote concentration relaxation curves drawn by using the power-law function kernel with $\alpha=0.2, 0.5$ and $0.8$ (from top to bottom). In this simulation the relaxation coefficient $\lambda=1.0$.}
 \label{figure3}
\end{figure}

Numerical solutions of the fractional derivative relaxation equation with a power-law or stretched exponential kernel are depicted in Figure \ref{figure3}. On the one hand, the fractional derivative relaxation model with a stretched exponential kernel exhibits much faster decay at small times and much slower decay at large times for a small order of fractional derivative (such as $\alpha=0.2$). On the other hand, the fractional derivative relaxation model with a stretched exponential kernel produces a faster decay than that with a power-law kernel for a large $\alpha$ (such as 0.8). This implies that the fractional derivative model with a stretched exponential kernel can describe a wider range of relaxation phenomena, compared to that with a power-law kernel.
% Figure 2

\begin{figure}[t]
\begin{center}
\setlength{\unitlength}{0.012500in}%
{\includegraphics[width=1.0\textwidth]{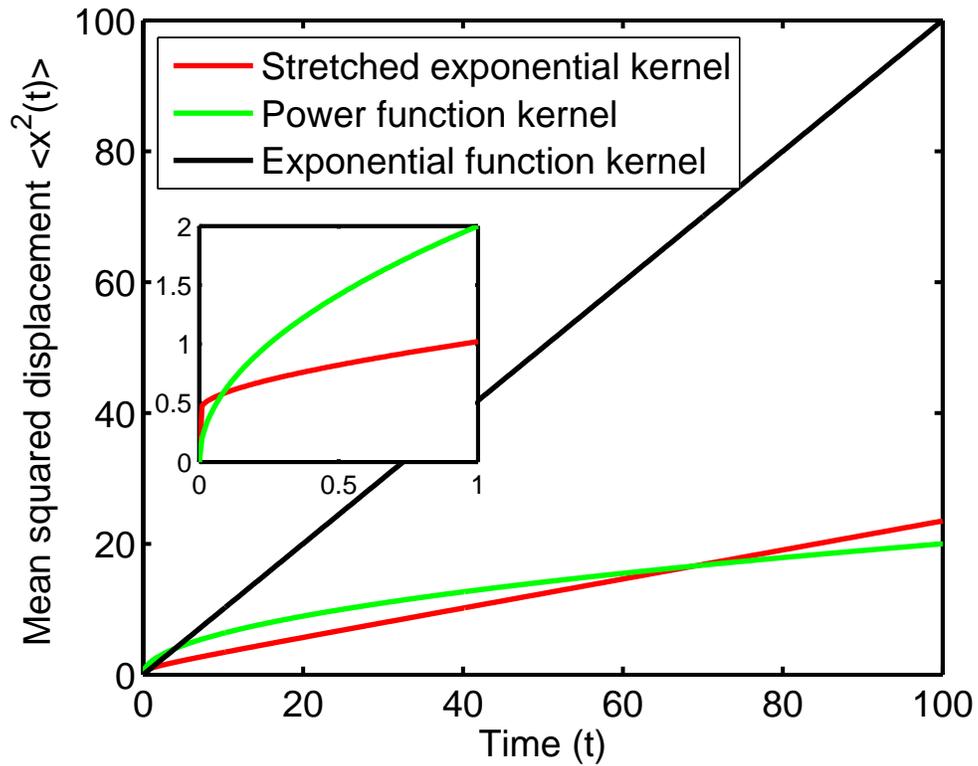}}\\
\end{center}
 \caption{The mean squared displacement (MSD) of the fractional derivative diffusion model with a stretched exponential function, power-law function, and exponential function as the kernel with $\alpha=0.5$ and the diffusion coefficient $K=1.0$.}
 \label{figure4}
\end{figure}

We further investigate the fractional derivative diffusion equation with a stretched exponential kernel. Figure \ref{figure4} shows truncated diffusive behavior of the fractional derivative diffusion equation with a stretched exponential kernel. This truncated diffusion behavior is analogous to the result of distributed-order or tempered fractional derivative models which may describe variations of MSD across scales \cite{Chechkin2002,Meerschaert2008}. A nonlinear increasing of the MSD which represents anomalous diffusion is observed at the initial time period, while an approximately linear increasing of the MSD with time is observed at the long time range. It should be noted that a fractional derivative model with a stretched exponential or power-law kernel produces slower diffusive motion than a model with an exponential kernel.
% Figure 3

Figure \ref{figure1} shows that the power-law kernel represents a strong memory in long-time history, compared to non-singular kernels. To characterize the strong memory in complex systems using the stretched exponential kernel, we propose a modified definition of the fractional derivative with a stretched exponential function kernel as follows:
\begin{eqnarray}
\begin{array}{c}
\displaystyle{^{SC}D^\alpha f(t)=\frac{1}{(1-\alpha)^{1/\alpha}\Gamma(1+\alpha)}\int_a^t f'(\tau) exp[-\frac{\alpha }{1-\alpha} (\frac{t-\tau}{t})^{\alpha}] d \tau \;.}
\end{array}
\label{eq6}
\end{eqnarray}

\begin{figure}[t]
\begin{center}
\setlength{\unitlength}{0.012500in}%
{\includegraphics[width=1.0\textwidth]{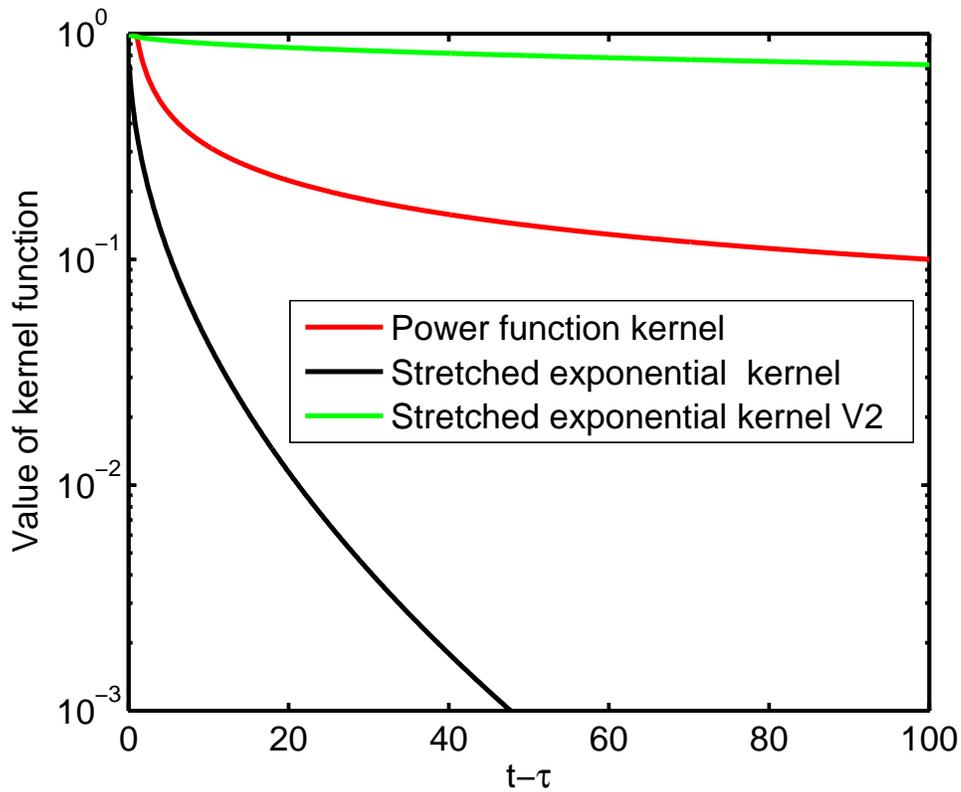}}\\
\end{center}
\caption{Memory property represented by different kernels with $\alpha=0.5$: the power-law function kernel $(t-\tau)^{-\alpha}$, the stretched exponential function kernel $exp[\frac{-\alpha*(t-\tau)^\alpha}{1-\alpha}]$, and the stretched exponential function(V2) kernel $exp[-\frac{\alpha}{1-\alpha}(\frac{t-\tau}{t})^\alpha]$.}
\label{figure5}
\end{figure}

Another advantage of definition (\ref{eq6}) is that the fractional dimension problem in the application of fractional derivative operators can be solved by introducing the expression of $(t-\tau)/t$. Figure \ref{figure5} shows that the stretched exponential kernel with $(t-\tau)/t$ exhibits super-slow decay of the system's memory, which may be a better tool to describe relaxation or diffusion processes with extremely strong memory.
% Figure 4

\section{Conclusions}

This study found that the fractional derivative with an exponential function kernel has limitations in characterizing the complex anomalous relaxation and diffusion processes. As a result, new fractional derivatives and integrals are required. An alternative notion of the fractional derivative with a stretched exponential function kernel is proposed to describe real-world complex relaxation and diffusion processes. In the limiting case we recover the exponential kernel.

There are still many issues which are worth further investigation. For examples, the lower limit of the integral in the new fractional derivative definition should be $-\infty$ from the mathematical viewpoint, otherwise the above analysis results are not strict.  The new definition therefore should be improved, and we will focus on this issue in a future study. Moreover, the applicability of new definition with stretched exponential function kernel, should be further verified by using a number of experimental data.

\section*{Acknowledgments}

This work was supported by the National Science Foundation of China under grants 11572112 and 11528205. Y. Zhang was partially supported by the National Science Foundation grant DMS-1460319 and the University of Alabama. This paper does not necessarily reflect the views of the funding agencies.

\newpage

%\section*{Figure introduction}
%
%\begin{figure}[t]

\end{document}